\documentclass[12pt]{article}
\usepackage{times}

\input{tcilatex}

\begin{document}

\title{\textbf{Exact solutions of the radial Schr\"{o}dinger equation for some
physical potentials }}
\author{Sameer M. Ikhdair\thanks{%
sikhdair@neu.edu.tr} \ and \ Ramazan Sever\thanks{%
sever@metu.edu.tr} \\
{\small \textsl{$^{\ast }$Department of Physics, \ Near East University,
Nicosia, North Cyprus, Mersin-10, Turkey }} \\
{\small \textsl{$^\dagger$Middle East Technical University,Department of
Physics,06531 Ankara, Turkiye }}}
\date{\today}
\maketitle

\begin{abstract}
By using an ansatz for the eigenfunction, we have obtained the exact
analytical solutions of the radial Schr\"{o}dinger equation for the
pseudoharmonic and Kratzer potentials in two dimensions. The energy levels
of all the bound states are easily calculated from this eigenfunction
ansatz. The normalized wavefunctions are also obtained.


PACS\ number: 03.65.-w; 03.65.Fd; 03.65.Ge.
\end{abstract}


\section{Introduction}

\noindent It is well known that the study of exactly solvable problems has
attracted much attention since the early study of non-relativistic quantum
mechanics. The exact solutions of the Schr\"{o}dinger equation for a
hydrogen atom (Coulombic) and for a harmonic oscillator in three dimensions
as well as in an arbitrary number of spatial $D$-dimensions represent two
typical examples in quantum mechanics [1-4]. The pseudoharmonic and Mie-type
potentials [5,6] are also two exactly solvable potentials other than the
Coulombic and anharmonic oscillator. In general, there are a few main
methods to study the exact solutions of quantum mechanical systems. The
first traditional method is to solve the second order differential
Schr\"{o}dinger equation by transforming it into some well known ordinary
differential equations, whose solutions are the special functions like the
confluent hypergeometric functions, associated Laguerre polynomials, the
Whittaker functions and others [2]. The second method is related with the
SUSYQM method, further closely with the factorization method [7]. The exact
quantization rule also shown its power in calculating the energy levels of
all the bound states for some exactly solvable quantum systems such as, the
Morse, the Rosen-Morse, the Kratzer, the harmonic oscillator, the first and
second P\"{o}schl-Teller and the pseudoharmonic oscillator potentials
[8-10]. The Nikiforov-Uvarov method [11] which is introduced for the
solution of Schr\"{o}dinger equation, to its energy levels, by transforming
it into hypergeometric type second order differential equations. The method
is based on the determination of the solution of wavefunction in terms of
special orthogonal functions for any general second-order differential
equation [12-15]. Also, the perturbative method introduced for calculating
the energy levels of some bound states of the Schr\"{o}dinger equation for
some approximately solvable quantum systems [16-19]. Further, the exact
solution of the $D$-dimensional Schr\"{o}dinger equation with some
anharmonic, pseudoharmonic and Kratzer potentials are obtained by applying
an ansatz to the wavefunction [20,21].

\"{O}z\c{c}elik and \c{S}im\c{s}ek have proposed an ansatz for eigenfunction
and presented the exact analytical solution of radial Schr\"{o}dinger
equation in three-dimensions of all the bound states with inverse-power
potentials [22]. Moreover, by applying this factorization ansatz to the
eigenfunction, Dong has carried out an exact analytical solutions of the
Schr\"{o}dinger equation in two-dimensions for some physical inverse-power
potentials [23].

It is well known in quantum mechanics, a total wave function provides
implicitly all relevant information about the behaviour of a physical
system. Hence if it is exactly solvable for a given potential, the wave
function can describe such a system completely. Until now, many efforts have
been made to solve the stationary Schr\"{o}dinger equation with anharmonic
potentials in three dimensions and two dimensions. With the same spirit, in
this work, we apply the method to study the solutions of the Schr\"{o}dinger
equation in two-dimensions for some pseudoharmonic and the modified Kratzer
potentials due to their applications in physics [5,6,8,10,21,24].

This paper is organized as follows. In Section \ref{WA}, we apply the
factorization ansatz to the eigenfunction to obtain an exact analytic
solution to the stationary radial Schr\"{o}dinger equation in two-dimensions
for the molecular pseudoharmonic and modified Kratzer-Fues potentials. We
also obtain the bound state eigensolutions of these molecular diatomic
potentials by making a suitable ansatz to every wave function. The
concluding remarks will be given in Section \ref{CR}.

\section{Wavefunction Ansatz}

\label{WA}Consider the two-dimensional stationary Schr\"{o}dinger equation
with a potential $V(\rho )$ depending only on the distance $\rho $ from the
origin [25]:

\begin{equation}
\left( -\frac{\hbar ^{2}}{2\mu }\nabla ^{2}+V(\rho )\right) \psi (\rho
,\varphi )=E\psi (\rho ,\varphi ),
\end{equation}
leads to the following equation for the radial part of the wave function
[25,26]: 
\begin{equation}
\left[ \frac{\partial ^{2}}{\partial \rho ^{2}}+\frac{1}{\rho }\frac{%
\partial }{\partial \rho }+\frac{1}{\rho ^{2}}\frac{\partial ^{2}}{\partial
\varphi ^{2}}+\frac{2\mu }{\hbar ^{2}}\left( E-V(\rho )\right) \right] \psi
(\rho ,\varphi )=0.
\end{equation}
On expressing the wave function as the product

\begin{equation}
\psi (\rho ,\varphi )=r^{-1/2}R_{m}(\rho )\exp (\pm im\varphi ),\text{ }%
m=0,1,2,\cdots ,
\end{equation}
the resulting differential equation for $R_{m}(\rho )$ reads

\begin{equation}
\left[ \frac{d^{2}}{d\rho ^{2}}+\frac{2\mu }{\hbar ^{2}}\left( E-V(\rho
)\right) -\frac{\left( m^{2}-1/4\right) }{\rho ^{2}}\right] R_{m}(\rho )=0,
\end{equation}
where $m$ and $E$ denotes the angular momentum and the energy spectra,
respectively. From the works [22,23], the radial wavefunction $R_{m}(\rho )$
has the following ansatz:

\begin{equation}
R_{m}(\rho )=f_{m}(\rho )\exp \left[ g(\rho )\right] ,
\end{equation}
where

\begin{equation}
f_{m}(\rho )=\left\{ 
\begin{array}{c}
1\text{ \ \ \ \ \ \ \ \ \ \ \ \ \ when }m=0, \\ 
\prod_{j=1}^{m}(\rho -\alpha _{j}^{(m)})\text{ \ when }m=1,2,3,\cdots
\end{array}
\right.
\end{equation}
and

\begin{equation}
g(\rho )=a\rho ^{2}+b\ln \rho ,\text{ }a<0.
\end{equation}
On inserting Eq.(5) into Eq.(4), one obtains

\begin{equation}
R_{m}^{\prime \prime }(\rho )-\left[ g^{\prime \prime }(\rho )+(g^{\prime
}(\rho ))^{2}+\left( \frac{f_{m}^{\prime \prime }(\rho )+2f_{m}^{\prime
}(\rho )g^{\prime }(\rho )}{f_{m}(\rho )}\right) \right] R_{m}(\rho )=0,
\end{equation}
where the prime denotes the derivative with respect to the variable $\rho .$
Following the Refs [27,28] it is found from Eq. (4) that $R_{m}^{\prime
\prime }(\rho )$ can be expressed as

\begin{equation}
-\frac{2\mu }{\hbar ^{2}}E+\frac{2\mu }{\hbar ^{2}}V(\rho )+\frac{\left(
m^{2}-1/4\right) }{\rho ^{2}}=g^{\prime \prime }(\rho )+(g^{\prime }(\rho
))^{2}+\frac{f_{m}^{\prime \prime }(\rho )+2f_{m}^{\prime }(\rho )g^{\prime
}(\rho )}{f_{m}(\rho )},
\end{equation}
which is the most fundamental equation for the following analysis.

\subsection{The pseudoharmonic potential}

This potential has been studied in three dimensions [5] using the polynomial
solution method and in $D$-dimensions using an ansatz for the wave function
[21]. It has the following form [5,21]: 
\begin{equation}
V(\rho )=D_{e}\left( \frac{\rho }{\rho _{e}}-\frac{\rho _{e}}{\rho }\right)
^{2},
\end{equation}
which can be simply rewritten in the form of isotropic harmonic oscillator
plus inverse quadratic potential [5,10,29] as 
\begin{equation}
V(r)=A\rho ^{2}+\frac{B}{\rho ^{2}}+C,\text{ }A,B>0,
\end{equation}
where $A=D_{e}\rho _{e}^{-2},$ $B=D_{e}\rho _{e}^{2}$ and $C=-2D_{e}.$

First of all, we study the ground state ($m=0$) so that we take $f_{0}(\rho
)=1$ and $g(\rho )$ in Eq. (7) to solve Eq. (9)

\begin{equation}
C^{\prime }-E^{\prime }+A^{\prime }\rho ^{2}+\frac{B^{\prime }+m^{2}-\frac{1%
}{4}}{\rho ^{2}}=2a\left( 1+2b\right) +4a^{2}\rho ^{2}+\frac{b\left(
b-1\right) }{\rho ^{2}},
\end{equation}
where $\ E^{\prime }=\frac{2\mu }{\hbar ^{2}}E,$ $A^{\prime }=\frac{2\mu }{%
\hbar ^{2}}A,$ $B^{\prime }=\frac{2\mu }{\hbar ^{2}}B$ and $C^{\prime }=%
\frac{2\mu }{\hbar ^{2}}C.$ \ On equating coefficients of $\rho ^{p},$ in
Eq. (12), for all $p=-2,0,2,$ the relations between the potential parameters
and the coefficients $a$ and $b$ are as follows: 
\begin{equation}
B^{\prime }-\frac{1}{4}=b^{2}-b,\text{ }A^{\prime }=4a^{2},\text{ }%
-E^{\prime }+C^{\prime }=2a(1+2b).
\end{equation}
Hence, we find from the first equation in (13) that $b=\frac{1}{2}+\sqrt{%
B^{\prime }}$ and from the second equation, $a=\pm \frac{\sqrt{A^{\prime }}}{%
2},$ but only the negative root yields a regular wave function at $\rho =0,$
and of course we choose $a=-\frac{\sqrt{A^{\prime }}}{2}$ so that $A^{\prime
}>0.$ Finally, from the third equation in (13), the energy is

\begin{equation}
E_{0}=C+\sqrt{\frac{2\hbar ^{2}A}{\mu }}\left( 1+\sqrt{\frac{2\mu B}{\hbar
^{2}}}\right) ,
\end{equation}
and the corresponding eigenfunction, by virtue of Eq. (3), can be written as

\begin{equation}
\psi _{0}(\rho ,\varphi )=N_{0}\exp \left[ -\frac{1}{2}\sqrt{\frac{2\mu A}{%
\hbar ^{2}}}\rho ^{2}\right] \rho ^{\sqrt{\frac{2\mu B}{\hbar ^{2}}}}.
\end{equation}
Secondly, for the first node $(m=1),$ we use $f_{1}(\rho )=\rho -\alpha
_{1}^{(1)}$ and $g(\rho )$ given in Eq. (7) to solve Eq. (9),

\begin{equation}
C^{\prime }-E^{\prime }+A^{\prime }\rho ^{2}+\frac{B^{\prime }+\frac{3}{4}}{%
\rho ^{2}}=2a\left( 3+2b\right) +4a^{2}\rho ^{2}+\frac{b\left( b+1\right) }{%
\rho ^{2}}.
\end{equation}
The relations between the potential parameters and the coefficients $a,b$
and $\alpha _{1}^{(1)}$ are

\begin{equation}
B^{\prime }+\frac{3}{4}=b^{2}+b,\text{ }A^{\prime }=4a^{2},-E^{\prime
}+C^{\prime }=2a(3+2b),\text{ }\alpha _{1}^{(1)}=0,
\end{equation}
which are solved as

\begin{equation}
b=-\frac{1}{2}+\sqrt{B^{\prime }+1},a=-\frac{1}{2}\sqrt{\text{ }A^{\prime }},%
\text{ }E_{1}^{\prime }=C^{\prime }+\sqrt{\text{ }A^{\prime }}(3+2b).
\end{equation}
So that the energy eigenvalue reads

\begin{equation}
E_{1}=C+\sqrt{\frac{2\hbar ^{2}A}{\mu }}\left( 1+\sqrt{\frac{2\mu B}{\hbar
^{2}}+1}\right) ,
\end{equation}
and the corresponding eigenfunction can be written as (cf. (3))

\begin{equation}
\psi _{1}(\rho ,\varphi )=N_{1}\exp \left[ -\frac{1}{2}\sqrt{\frac{2\mu A}{%
\hbar ^{2}}}\rho ^{2}\right] \rho ^{\sqrt{\frac{2\mu B}{\hbar ^{2}}+1}%
}e^{\pm i\varphi }.
\end{equation}
We follow the same steps for the second node ($m=2)$ (i.e. $f_{2}(\rho
)=(\rho -\alpha _{1}^{(2)})(\rho -\alpha _{2}^{(2)})$) and $g(\rho )$ is
defined in Eq.(7) which imply that the following algebraic equation holds

\begin{equation}
C^{\prime }-E^{\prime }+A^{\prime }\rho ^{2}+\frac{B^{\prime }+\frac{15}{4}}{%
\rho ^{2}}=2a\left( 5+2b\right) +4a^{2}\rho ^{2}+\frac{(b+1)\left(
b+2\right) }{\rho ^{2}},
\end{equation}
which leads to equations

\begin{equation}
B^{\prime }+\frac{15}{4}=(b+1)\left( b+2\right) ,\text{ }A^{\prime }=4a^{2},%
\text{ }-E^{\prime }+C^{\prime }=2a(5+2b),\alpha _{1}^{(2)}=\alpha
_{2}^{(2)}=0.
\end{equation}
Therefore, the energy eigenvalue is 
\begin{equation}
E_{2}=C+\sqrt{\frac{2\hbar ^{2}A}{\mu }}\left( 1+\sqrt{\frac{2\mu B}{\hbar
^{2}}+4}\right) ,
\end{equation}
and the eigenfunction reads

\begin{equation}
\psi _{2}(\rho ,\varphi )=N_{2}\exp \left[ -\frac{1}{2}\sqrt{\frac{2\mu A}{%
\hbar ^{2}}}\rho ^{2}\right] \rho ^{\sqrt{\frac{2\mu B}{\hbar ^{2}}+4}%
}e^{\pm i2\varphi }.
\end{equation}
It is now possible to understand the general structure of the calculation
with arbitrary value of the angular momentum $m$. One has the general
algebraic equation

\begin{equation}
C^{\prime }-E^{\prime }+A^{\prime }\rho ^{2}+\frac{B^{\prime }+m^{2}-\frac{1%
}{4}}{\rho ^{2}}=2a\left( 2m+2b+1\right) +4a^{2}\rho ^{2}+\frac{(b+m)\left(
b+m-1\right) }{\rho ^{2}},
\end{equation}
which yields the equations:

\begin{equation}
B^{\prime }+m^{2}-\frac{1}{4}=(b+m)\left( b+m-1\right) ,\text{ }A^{\prime
}=4a^{2},\text{ }-E^{\prime }+C^{\prime }=2a(2m+2b+1).
\end{equation}
Finally, the general formula for energy eigenvalue reads

\begin{equation}
E_{m}=C+\sqrt{\frac{2\hbar ^{2}A}{\mu }}\left( 1+\sqrt{\frac{2\mu B}{\hbar
^{2}}+m^{2}}\right) ,\text{ }m=0,1,2,\cdots
\end{equation}
and the corresponding normalized radial wave function is factorized in the
form

\begin{equation}
\psi _{m}(\rho ,\varphi )=N\exp \left[ -\frac{1}{2}\sqrt{\frac{2\mu A}{\hbar
^{2}}}r^{2}\right] r^{\sqrt{\frac{2\mu B}{\hbar ^{2}}+m^{2}}}e^{\pm
im\varphi },\text{ }m=0,1,2,\cdots
\end{equation}
where all normalization constants can be evaluated from the condition

\begin{equation}
\int\limits_{0}^{\infty }\left| \psi ^{(n)}(r)\right| ^{2}rdr=1.
\end{equation}
For example, the explicit calculation shows that

\begin{equation}
N=\left[ \frac{\left( 2\sqrt{\frac{2\mu A}{\hbar ^{2}}}\right) ^{1+\sqrt{%
\frac{2\mu B}{\hbar ^{2}}+m^{2}}}}{\left( \sqrt{\frac{2\mu B}{\hbar ^{2}}%
+m^{2}}\right) !}\right] ^{1/2},
\end{equation}
which is the normalization constant.

\subsection{The Mie-type potentials}

This potential has been studied in the $D$ dimensions using the polynomial
solution and the ansatz wave function method [6,21]. An example on this type
of potentials is the standard Morse [30] or Kratzer-Fues [31,32] potential
of the form [6,21,24,33]

\begin{equation}
V(\rho )=-D_{e}\left( \frac{2\rho _{e}}{\rho }-\frac{\rho _{e}^{2}}{\rho ^{2}%
}\right) ,
\end{equation}
where $D_{e}$ is the dissociation energy between two atoms in a solid and $%
\rho _{e}$ is the equilibrium internuclear seperation. The standard Kratzer
potential is modified by adding a $D_{e}$ term to the potential. A new type
of this potential is the modified Kratzer-type of molecular potential [21,24]

\begin{equation}
V(\rho )=D_{e}\left( \frac{\rho -\rho _{e}}{\rho }\right) ^{2},
\end{equation}
and hence it is shifted in amount of $D_{e}.$ The potential in Eq. (32) has
been studied in $D$ dimensions [21] by making the wave function ansatz [20].
However, this potential has also been discussed before in three dimensions
[24] and in $D$ dimensions [6]. This potential [33] can be simply taken as 
\begin{equation}
V(r)=\frac{A}{\rho }+\frac{B}{\rho ^{2}}+C,
\end{equation}
where $A=-D_{e}r_{e},$ $B=D_{e}r_{e}^{2}$ and $C=D_{e}$ [21,24]$.$

First of all, for the ground state ($m=0$), we take $f_{0}(\rho )=1$ and $%
g(\rho )$ in Eq. (7) to solve Eq. (9)

\begin{equation}
C^{\prime }-E^{\prime }+\frac{A^{\prime }}{\rho }+\frac{B^{\prime }+m^{2}-%
\frac{1}{4}}{\rho ^{2}}=a^{2}+\frac{2ab}{\rho }+\frac{b\left( b-1\right) }{%
\rho ^{2}},
\end{equation}
where $\ E^{\prime }=\frac{2\mu }{\hbar ^{2}}E,$ $A^{\prime }=\frac{2\mu }{%
\hbar ^{2}}A,$ $B^{\prime }=\frac{2\mu }{\hbar ^{2}}B$ and $C^{\prime }=%
\frac{2\mu }{\hbar ^{2}}C.$ \ On equating coefficients of $\rho ^{p},$ for
all $p=-2,0,2,$ the relations between the potential parameters and the
coefficients $a$ and $b$ are

\begin{equation}
B^{\prime }-\frac{1}{4}=b(b-1),\text{ }A^{\prime }=2ab,\text{ }-E^{\prime
}+C^{\prime }=a^{2}.
\end{equation}
Hence, we find from the first equation in (35) that $b=\frac{1}{2}+\sqrt{%
B^{\prime }}$ and from the second equation in (35) is solved by $a=\frac{%
A^{\prime }}{1+2\sqrt{B^{\prime }}}.$ Finally, the energy eigenvalue reads

\begin{equation}
E_{0}=C-\frac{2\mu A^{2}/\hbar ^{2}}{\left[ 1+2\sqrt{\frac{2\mu B}{\hbar ^{2}%
}}\right] ^{2}},
\end{equation}
and the normalized wave function is

\begin{equation}
\psi _{0}(\rho ,\varphi )=N_{0}\exp \left[ -\sqrt{-\frac{2\mu }{\hbar ^{2}}%
\left( E-C\right) }\rho \right] \rho ^{\sqrt{\frac{2\mu B}{\hbar ^{2}}}}.
\end{equation}
Secondly, for the first node $(m=1),$ we use $f_{1}(\rho )=\rho -\alpha
_{1}^{(1)}$ and $g(\rho )$ given in Eq. (7) to solve Eq. (9),

\begin{equation}
C^{\prime }-E^{\prime }+\frac{A^{\prime }}{\rho }+\frac{B^{\prime }+\frac{3}{%
4}}{\rho ^{2}}=a^{2}+\frac{2a(1+b)}{\rho }+\frac{b\left( b+1\right) }{\rho
^{2}}.
\end{equation}
Hence, the relations between the potential parameters and the coefficients $%
a,b$ and $\alpha _{1}^{(1)}$ are

\begin{equation}
B^{\prime }+\frac{3}{4}=b^{2}+b,\text{ }A^{\prime }=2a(1+b),\text{ }%
-E^{\prime }+C^{\prime }=a^{2},\text{ }\alpha _{1}^{(1)}=0,
\end{equation}
and so the exact energy eigenvalue is

\begin{equation}
E_{1}=C-\frac{2\mu A^{2}/\hbar ^{2}}{\left[ 1+2\sqrt{\frac{2\mu B}{\hbar ^{2}%
}+1}\right] ^{2}},
\end{equation}
and the corresponding wave function is

\begin{equation}
\psi _{1}(\rho ,\varphi )=N_{1}\exp \left[ -\sqrt{-\frac{2\mu }{\hbar ^{2}}%
\left( E-C\right) }\rho \right] \rho ^{\sqrt{\frac{2\mu B}{\hbar ^{2}}+1}%
}e^{\pm i\varphi }.
\end{equation}
We follow the same steps for the second node, $m=2$ (i.e. $f_{2}(\rho
)=(\rho -\alpha _{1}^{(2)})(\rho -\alpha _{2}^{(2)})$) and $g(\rho )$ is
defined in Eq.(7) to find

\begin{equation}
C^{\prime }-E^{\prime }+\frac{A^{\prime }}{\rho }+\frac{B^{\prime }+\frac{15%
}{4}}{\rho ^{2}}=a^{2}+\frac{2a(b+2)}{\rho }+\frac{(b+1)\left( b+2\right) }{%
\rho ^{2}},
\end{equation}
and their solutions are hence found to be

\begin{equation}
B^{\prime }+\frac{15}{4}=(b+1)\left( b+2\right) ,\text{ }A^{\prime }=2a(b+2),%
\text{ }-E^{\prime }+C^{\prime }=a^{2},\text{ }\alpha _{1}^{(2)}=\alpha
_{2}^{(2)}=0.
\end{equation}
Therefore, the energy eigenvalue is

\begin{equation}
E_{2}=C-\frac{2\mu A^{2}/\hbar ^{2}}{\left[ 1+2\sqrt{\frac{2\mu B}{\hbar ^{2}%
}+4}\right] ^{2}},
\end{equation}
and hence the corresponding wave function is

\begin{equation}
\psi _{2}(\rho ,\varphi )=N_{2}\exp \left[ -\sqrt{-\frac{2\mu }{\hbar ^{2}}%
\left( E-C\right) }\rho \right] \rho ^{\sqrt{\frac{2\mu B}{\hbar ^{2}}+4}%
}e^{\pm i2\varphi }.
\end{equation}
Thus, in general, for any arbitrary $m,$ we find that

\begin{equation}
C^{\prime }-E^{\prime }+\frac{A^{\prime }}{\rho }+\frac{B^{\prime }+m^{2}-%
\frac{1}{4}}{\rho ^{2}}=a^{2}+\frac{2a(b+m)}{\rho }+\frac{(b+m)\left(
b+m-1\right) }{\rho ^{2}},
\end{equation}
which leads to the following algebraic equations

\begin{equation}
B^{\prime }+m^{2}-\frac{1}{4}=(b+m)\left( b+m-1\right) ,\text{ }A^{\prime
}=2a(b+m),\text{ }-E^{\prime }+C^{\prime }=a^{2}.
\end{equation}
Finally, the energy eigenvalue is

\begin{equation}
E_{m}=C-\frac{2\mu A^{2}/\hbar ^{2}}{\left[ 1+2\sqrt{\frac{2\mu B}{\hbar ^{2}%
}+m^{2}}\right] ^{2}},
\end{equation}
and the corresponding normalized wave function is

\begin{equation}
\psi _{m}(\rho ,\varphi )=N\exp \left[ -\sqrt{-\frac{2\mu }{\hbar ^{2}}%
\left( E-C\right) }\rho \right] \rho ^{\sqrt{\frac{2\mu B}{\hbar ^{2}}+m^{2}}%
}e^{\pm im\varphi },
\end{equation}
where the normalization constant

\begin{equation}
N=\frac{\left( 2\sqrt{-\frac{2\mu }{\hbar ^{2}}(E-C)}\right) ^{1+\sqrt{\frac{%
2\mu B}{\hbar ^{2}}+m^{2}}}}{\sqrt{\left( 2\sqrt{\frac{2\mu B}{\hbar ^{2}}%
+m^{2}}+1\right) !}}.
\end{equation}

\section{Concluding Remarks}

\label{CR}We have easily obtained the exact bound state solutions of the
two-dimensional radial Schr\"{o}dinger equation for two general potential
forms representing the pseudoharmonic [5,21] and modified Kratzer molecular
[6,21] potentials by using the wave function ansatz method [22,23]. The
presented procedure in this study is systematical and efficient in finding
the exact energy spectra and corresponding wave functions of the
Schr\"{o}dinger equation for any desired quantum system. This method is
simple in producing the exact bound state solution for further anharmonic
potentials.

\section{Acknowledgments}

This research was partially supported by the Scientific and Technological
Research Council of Turkey. S.M. Ikhdair is grateful to his family members
for their assistance and love.

\end{document}